\input harvmac
\input epsf

\noblackbox


\def\bfone{\relax{\rm 1\kern-.35em 1}}
\def\inbar{\vrule height1.5ex width.4pt depth0pt}

\def\IC{\relax\,\hbox{$\inbar\kern-.3em{\rm C}$}}
\def\ID{\relax{\rm I\kern-.18em D}}
\def\IF{\relax{\rm I\kern-.18em F}}
\def\IH{\relax{\rm I\kern-.18em H}}
\def\II{\relax{\rm I\kern-.17em I}}
\def\IN{\relax{\rm I\kern-.18em N}}
\def\IP{\relax{\rm I\kern-.18em P}}
\def\IQ{\relax\,\hbox{$\inbar\kern-.3em{\rm Q}$}}
\def\us#1{\underline{#1}}
\def\IR{\relax{\rm I\kern-.18em R}}
\font\cmss=cmss10 \font\cmsss=cmss10 at 7pt
\def\ZZ{\relax\ifmmode\mathchoice
{\hbox{\cmss Z\kern-.4em Z}}{\hbox{\cmss Z\kern-.4em Z}}
{\lower.9pt\hbox{\cmsss Z\kern-.4em Z}}
{\lower1.2pt\hbox{\cmsss Z\kern-.4em Z}}\else{\cmss Z\kern-.4em
Z}\fi}
\def\a{\alpha}  
 
 \def\la{\lambda}
 \def\s{\sigma}
\def\th{\theta}

\def\nup#1({Nucl.\ Phys.\ $\us {B#1}$\ (}
\def\plt#1({Phys.\ Lett.\ $\us  {B#1}$\ (}
\def\cmp#1({Comm.\ Math.\ Phys.\ $\us  {#1}$\ (}
\def\prp#1({Phys.\ Rep.\ $\us  {#1}$\ (}
\def\prl#1({Phys.\ Rev.\ Lett.\ $\us  {#1}$\ (}
\def\prv#1({Phys.\ Rev.\ $\us  {#1}$\ (}
\def\mpl#1({Mod.\ Phys.\ Let.\ $\us  {A#1}$\ (}
\def\ijmp#1({Int.\ J.\ Mod.\ Phys.\ $\us{A#1}$\ (}
\def\jag#1({Jour.\ Alg.\ Geom.\ $\us {#1}$\ (}
\def\tit#1|{{\it #1},\ }

\def\Coe#1.#2.{{#1\over #2}}

\def\coe#1.#2.{\relax{\textstyle {#1 \over #2}}\displaystyle}
\def\half{{1 \over 2}}

\def\sqgn{\sqrt{gN}}
\def\tf{\widetilde f}
\def\th{\widetilde h}

%
\lref\JMalda{J. Maldacena, {\it The large N limit of superconformal
field theories and supergravity}, hep-th/9711200.}
\lref\IMSY{}
\lref\MDJR{M.J.~Duff and J.~Rahmfeld  \nup{481} (1996) 332, 
hep-th/9605085.}
\lref\KStelle{K.S.~Stelle  {\it BPS Branes in Supergravity,} 
in Proceedings of the ICTP Summer School in High-energy Physics 
and Cosmology, Trieste, Italy, 10 Jun -- 26 Jul 1996 and 2 Jun -- 
11 Jul 1997, hep-th/9803116.}
\lref\JMaldb{J. Maldacena, {\it Wilson loops in large N field 
theories}, HUTP-98/A014, hep-th/9803002.}
\lref\RY{S.-J. Rey and J.~Yee, {\it Macroscopic strings as heavy quarks in 
large {N} gauge theory and anti-de Sitter supergravity,} hep-th/9803001.}
\lref\RTY{S.-J. Rey, S.~Theisen and J.~Yee, {\it Wilson-Polyakov Loop at 
Finite Temperature in Large $N$ Gauge Theory and
     Anti-de Sitter Supergravity,} hep-th/9803135.}
\lref\BISY{A.~Brandhuber, N.~Itzhaki, J.~Sonnenschein and S.~Yankielowicz,
{\it Wilson Loops in the Large $N$ Limit at Finite Temperature}, 
hep-th/9803137}
\lref\BISYII{A.~Brandhuber, N.~Itzhaki, J.~Sonnenschein and S.~Yankielowicz,
{\it Wilson Loops, Confinement, and Phase Transitions in Large $N$ Gauge 
Theories from Supergravity}, hep-th/9803263.}
\lref\ML{M.~Li, {\it 't Hooft Vortices on D-branes}, hep-th/9803252;
 {\it 't Hooft vortices and phases of large $N$ gauge theory},
hep-th/9804175.}
\lref\DP{ U.~Danielsson and  A.~Polychronakos, {\it Quarks, monopoles and 
dyons at large $N$,} hep-th/9804141.}
\lref\JMin{J.~Minahan, {\it Quark-Monopole Potentials in Large $N$ Super 
Yang-Mills}, hep-th/9803111.}
\lref\LanLif{L.D.~Landau and E.M.~Lifshitz, {\it Mechanics,}
Third Edition, Pergamon Press, (1973).}
\lref\Wittherm{E.~Witten, {Anti-de Sitter Space, Thermal Phase Transition, 
And Confinement In Gauge Theories}, hep-th/9803131.}
\lref\MW{J. Minahan and N. Warner}
\lref\Arvis{J.~Arvis, {\it Phys. Lett.} {\bf 127B} (1983) 106.}
\lref\Alv{O.~Alvarez, {\it Phys. Rev. D} {\bf 24} (1981) 440.}
\lref\Luscher{M.~L\"uscher, {\it Nucl. Phys.} {\bf B180} (1981) 317.}
\lref\HashOz{A.~Hashimoto and Y.~Oz, {\it Aspects of QCD Dynamics from String 
Theory}, hep-th/9809106.}
\lref\KJMN{R.de Mello Koch, A.Jevicki, M.Mihailescu and J.P.Nunes, 
Phys.Rev. {\bf D58} (1998) 105009, hep-th/9806125}
\lref\KT{I. Klebanov and A. Tseytlin, {\it D-Branes and Dual Gauge Theories 
in Type 0 Strings}, hep-th/9811035.}
\lref\BG{O. Bergman and M. Gaberdiel, {\it Nucl. Phys.} {\bf B499} (1997) 183,
hep-th/9701137.}
\lref\Horava{P. Horava, {\it On QCD String Theory and AdS Dynamics},
hep-th/9811028.}
\lref\Polyakov{A. Polyakov, {\it Nucl. Phys.} {\bf B268} (1986) 406.}
\lref\Zyskin{M. Zyskin, {\it A note on the glueball mass spectrum}, 
hep-th/9806128.}
\lref\PP{A. Peet, J. Polchinski, {\it UV/IR Relations in AdS Dynamics},
hep-th/9809022.}
\lref\ABKS{O. Aharony, M. Berkooz, D. Kutasov and N. Seiberg,
{\it Linear Dilatons, NS5-branes and Holography}, hep-th/9808149.}
\lref\GO{D. Gross and H. Ooguri, {\it Phys. Rev.} {\bf D58} (1998) 106002,
hep-th/9805129.}
\lref\COOT{C. Csaki, H. Ooguri, Y. Oz and J. Terning, {\it Glueball Mass 
Spectrum From Supergravity}, hep-th/9806021.}
\lref\CORT{C. Csaki, Y. Oz, J. Russo and J. Terning, {\it Large $N$ QCD from 
Rotating Branes}, hep-th/9810186.}
\lref\Russo{J. Russo, {\it  New Compactifications of Supergravities and Large 
$N$ QCD},  hep-th/9808117.}
\lref\GKP{S. Gubser, I. Klebanov and  A. Polyakov, {\it Phys. Lett.} 
{\bf B428} (1998) 105, hep-th/9802109.}
\lref\Witten{E. Witten, {\it Adv. Theor. Math. Phys.} {\bf 2} (1998) 253,
hep-th/9802150.}
\lref\DH{L. Dixon and J. Harvey, {\it Nucl. Phys.} {\bf B274} (1986) 93.}
\lref\SW{N. Seiberg and E. Witten, {\it Nucl. Phys.} {\bf B276} (1986) 272.}
\lref\Gross{D. Gross, Strings '98 talk.}
\lref\GKPeet{S. Gubser, I. Klebanov and A. Peet, {\it Phys. Rev.} {\bf D54}
 (1996) 3915, hep-th/9602135.}
\lref\Cave{A. Polyakov, {\it The Wall of the Cave}, /hep-th/9809057}
%
%
\Title{\vbox{
\hbox{CALT-68-2203}
\hbox{\tt hep-th/9811156}
}}{\vbox{\centerline{\hbox{ Glueball Mass Spectra and Other Issues for 
}}
\vskip 8 pt
\centerline{ \hbox{Supergravity Duals of QCD Models}}}}
\centerline{Joseph A. Minahan}
\bigskip
\centerline{{\it California Institute of Technology,
Pasadena, CA 91125, USA}}
\bigskip

We derive WKB expressions for glueball masses of various 
finite temperature supergravity
models.  The results are very close to recent numerical computations.
We argue that the spectra has some universality that depends only
on the dimension of the {\it AdS} space and the singularity structure of
the horizon.  
This explains the stability of the $0^{++}$ glueball mass ratios between
various models.  We also consider the recently proposed 
nonsupersymmetric model arising
from the type $0$ string.  In the supergravity limit of this model, 
the heavy quark potential
has an effective coupling with $1/(\log u)$ behavior in the {\it UV}.  
Unfortunately, the supergravity solution implies that  the heavy quark 
potential is still coulombic in the infrared,
with an  effective coupling of order 1.  We also argue that the type 0
  supergravity background solution
 does not have normalizable glueball solutions.

\vskip .3in

\Date{\sl {November, 1998}}
\vfil
\eject
\parskip=4pt plus 15pt minus 1pt
\baselineskip=15pt plus 2pt minus 1pt
%
\newsec{Introduction}

One of the many interesting developments to arise out of Maldacena's 
conjecture \JMalda\refs{\GKP,\Witten} is the ability to study 
nonsupersymmetric large $N$ gauge
theories at strong coupling \refs{\Wittherm\RTY\BISYII\ML\GO{--}\COOT}.  One
studies a $d$ dimensional euclidean gauge theory at finite temperature, 
which is equivalent to a theory with $d-1$ noncompact directions and a 
Euclidean time compactified on a circle of circumference $\beta$.  As
was pointed out by Witten \Wittherm\ the Maldacena conjecture relates
wave equations in an $AdS$ blackhole background to  two point correlation
functions of a finite temperature Yang-Mills gauge theory. 

Using this conjecture Witten argued that the dilaton wave equation in this
background implies a discrete glueball spectrum with a finite gap.  
This spectrum
was studied more closely in \COOT\ and also in \refs{\KJMN,\Zyskin} where
comparisons were made between the supergravity results and lattice
gauge theory results.  However, the strong coupling behavior of QCD is highly 
nonuniversal, so there really is no reason to expect much similarity
between the lattice results and the supergravity results, beyond the fact that
the spectra for both theories is discrete with a finite gap.  

However, one might hope to find some universality within different 
supergravity models.  In particular, other supergravity models were recently
studied that correspond to finite
temperature QCD with its $R$ symmetry group broken\refs{\Russo,\CORT}.
This has the nice feature of getting rid of some of the unwanted
Kaluza-Klein states.  It was noted in \CORT\ that there
seemed to be some universality in the mass ratios of $J^{PC}=0^{++}$ glueball
states for the different supergravity models.  One of the purposes of
this paper is to explain this universality by finding WKB approximations for
the glueball spectra.  As it turns out, the leading order term depends on
the particular supergravity theory being considered.  However, the subleading
term has universal behavior, depending only on the dimension of the $AdS$ space
and the horizon singularity. 
One could then speculate that if ``real'' QCD has a supergravity
dual that is asymptotically $AdS_n$, then it too will have some universal
behavior, that depends only on $n$ and the singularity structure at 
a horizon.

In all of the above models, supersymmetry was broken by turning on a 
temperature.   Therefore, supersymmetry is restored in the ultraviolet
and hence the supergravity solutions do not exhibit asymptotic freedom.
 Recently, a nonsupersymmetric gauge model was proposed that
arises from D3 branes in the nonsupersymmetric type 0 theory \refs{\Cave,\KT}. 
Since
supersymmetry is never restored, one should expect to see running of
the coupling in the {\it UV}.  In principle, one
could also apply the WKB analyis to this model and derive its glueball
 spectrum,  up to an overall scale.   

The type 0 theory
 has a tachyon in the bulk that presumably gets an expectation value.
The authors in \KT\ derived a supergravity action for this model.  We will
find asymptotic solutions to the equations of motion coming from this 
supergravity action. 
The results found here should be taken with a grain of salt since the
background metric has curvature that is either greater than or
roughly equal to  the string scale in both the {\it UV} and {\it IR}.
Nonetheless, one still hopes that the supergravity results are qualitatively
correct, as in the case for the entropy of ${\cal N}=4$ super Yang-Mills
at finite temperature, where the supergravity result differed from the 
perturbative Yang-Mills result by a factor of $3/4$ \GKPeet. 
In fact, in the type 0 case we do
 find a running of the effective coupling in the {\it UV}.
However, in the infrared we do not find confinement nor do we find 
normalizable glueball solutions.  This suggests that one must consider
the full $\s$-model to see such behavior.

In section 2 we derive WKB expressions for the masses of $0^{++}$ in the 
finite temperature models described in \Wittherm.  We compare these results
to the recent numerical results and we find excellent agreement.  We also
describe the six dimensional model with five uncompactified directions and
argue that the glueball spectrum has a finite gap with a continuous spectrum
above the gap.  In section 3 we derive WKB expressions for more general
supergravity models, and we show that these expressions have a general form
that depends on the dimension of the asymptotic $AdS$ space and the 
singularity at the horizon.  This explains the recently noted stability
of the spectrum for the class of models discussed in \CORT.  
Using this analysis we also find WKB 
expressions
 for the $0^{--}$ and $0^{-+}$ glueballs in 3 and 4 dimensions
respectively.  We again find good agreement with the numerical results.
In section 4 we describe our findings for the type 0 model.  We derive
expressions for the metric and the coupling in the ultraviolet and
infrared and use this
to find the heavy quark potential in these two limits.  We also argue that
the {\it IR} behavior of the metric and coupling does not allow 
for normalizable
glueball solutions.  In section 5 we present our conclusions.

\newsec{WKB Masses for $0^{++}$ Glueballs}

In this section we compute masses using a WKB approximation for the
$0^{++}$ glueballs in the supergravity models in \Wittherm. In the next
section we will consider more general cases as well as the WKB solutions
for the other glueballs recently discussed in the literature.

The $0^{++}$ glueball spectrum is governed by the dilaton wave equation
in the appropriate background.  Consider first the background arising
fron $N$ stationary D$p$ branes at  finite temperature $T$.  The metric
in the near horizon limit is
\eqn\metricft{\eqalign{
ds^2~=~\a'\Biggl[{U^{(7-p)/2}\over\sqgn}&\left(
\left(1-\left({U_T\over U}\right)^{7-p}\right)dt^2+ dx_{i}^2\right)\cr
&
~+~{\sqgn\over U^{(7-p)/2}}\left({dU^2\over1-({U_T\over U})^{7-p}}~+~
U^2d\Omega_3^2\right)\Biggr],
}}
and with a dilaton background
\eqn\dilbg{
e^\phi~=~g\left({U^{7-p}\over gN}\right)^{(p-3)/4}.
}
The temperature $T$ is related to $U_T$ and the coupling by
\eqn\Temprel{
T~=~{7-p\over4\pi}~{U_T^{(5-p)/2}\over\sqgn}
}

The dilaton equation of motion is
\eqn\dileqm{
\partial_\mu e^{-2\phi} \sqrt{g}g^{\mu\nu}\partial_\nu\phi=0.
}
Assuming that $\phi$ is of the form $\phi=e^{ik\cdot x}\rho(u)$ with 
$k^2=-M^2$,
\dileqm\ reduces to 
\eqn\dileqmr{
\partial_U\left(U^{7-p}-U_T^{7-p}\right)U\partial_U\phi~+~M^2gNU\phi~=~0.
}
Defining a new variable $x=U^2$ and rescaling, \dileqmr\ reduces further to
\eqn\diffeq{
\partial (x^{2+1/n}-x)\partial\phi~+~\la\phi~=~0
}
where $n=2/(5-p)$ and
\eqn\larel{
\la=M^2\left({n+1\over4\pi nT}\right)^2.
}
The differential equation in \diffeq\ has singularities at $x=0,\infty$
and at all $1/(1+n)$ roots of unity, so solutions to this equation are
unknown for finite $n$.  In order to do the WKB approximation, we
define a new function $\psi=\sqrt{x-1\over x^{2+1/n}-x}\phi$, and we change
variables to $x=1+e^y$.  The equation in \diffeq\ now takes the form
\eqn\diffnew{
\psi''~+\left({\la\over f}e^y-\ha {f''\over f}+
{1\over4}\left({f'\over f}\right)^2\right)\psi~=~0,
}
where the primes denote derivatives with respect to $y$ and 
$f=(1-e^{-y})((e^y+1)^{1+1/n}-1)$.  

For large negative $y$, the term in
front of $\psi$ in \diffnew\ is approximately
\eqn\negapprox{
{\la\over f}e^y-\ha {f''\over f}+{1\over4}\left({f'\over f}\right)^2~\approx~
\left({\la n\over1+n}-{1\over4}(2+1/n)\right)e^y\qquad\qquad y<<0.
}
For large positive $y$ the asymptotic behavior for this term is
\eqn\posapprox{
{\la\over f}e^y-\ha {f''\over f}+{1\over4}\left({f'\over f}\right)^2~\approx~
\la e^{-y/n}-{(n+1)^2\over4n^2}\qquad\qquad y>>0.
}
Thus, for $\la$ sufficiently large
there will be two turning points at $y=-\infty$ and $y=y_0$, where
\eqn\turnpt{
y_0\approx n\log(4n\la/(n+1)^2).
}
Hence, the WKB approximation for
this curve gives
\eqn\WKBeq{
(m+1/2)\pi~=~\int_{-\infty}^{y_0}dy
\sqrt{{\la\over f}e^y-\ha {f''\over f}+{1\over4}\left({f'\over f}\right)^2},
\qquad\qquad m\ge0.
}
To leading order in $M$ we may approximate the WKB integral as
\eqn\WKBlead{
\int_{-\infty}^\infty dy \sqrt{\la^{1/2}}e^{y/2}f^{-1/2}~=~
\int_1^\infty dx{\sqrt{\la^{1/2}}\over\sqrt{x^{2+1/n}-x}}~=~
{M\over T}{\Gamma\left({1\over2+2n}\right)\over4\pi^{1/2}
\Gamma\left({2+n\over2+2n}\right)}.
}

Let us now consider the next order term in the $1/M$ expansion of \WKBeq. There
are two contributions to this constant piece.  There is one contribution 
because \WKBlead\ was integrated to $\infty$ instead of $y_0$.  Hence we should
subtract from
\WKBlead\ the term
\eqn\corrone{
\int_{y_0}^\infty dy \sqrt{\la e^y\over f}~=~ (n+1)~+~{\rm O}(1/\sqrt{\la}).
}
The other contribution comes from integrating the integrand in
\WKBeq\ near $y_0$.
Subtracting off the leading order term, this contribution is given by
\eqn\corrtwo{\eqalign{
&\int_{-\infty}^{y_0}dy
\sqrt{{\la\over f}e^y-\ha {f''\over f}+{1\over4}\left({f'\over f}\right)^2}-
\sqrt{\la e^y\over f}~\approx~\cr
&
\int_1^{e^{y_0}}{dx\over x^{1+1/(2n)}}
\left(\sqrt{\la-{(1+1/n)^2\over4} x^{1/n}}-
\sqrt{\la}\right)~=~\left(1-{\pi\over2}\right)(n+1)~+~{\rm O}(1/\sqrt{\la}).
}}
We have used the fact that  $f''/f$ and $f'/f$ are almost constant 
near the turning point $y=y_0$ for large $\sqrt{\la}$.  Hence, using
\larel, \WKBlead, \corrone\ and \corrtwo\ in \WKBeq, we find that
\eqn\WKBfin{
M^2~=~16\pi^3\left({\Gamma\left({2+n\over 2+2n}\right)\over
\Gamma\left({1\over 2+2n}\right)}\right)^2T^2 m(m+n)~+~{\rm O}(m^0)\qquad
m\ge1
.}

In the $QCD_3$ case, $n=1$ and hence the mass relation is
\eqn\WKBIII{\eqalign{
M^2~&=~8\pi \left(\Gamma\left({3\over 4}\right)\right)^4T^2 m(m+1)
~+~{\rm O}(m^0)\qquad m\ge1\cr
~&\approx~5.74216~(\pi T)^2m(m+1)
.}}
We have factored out a $\pi^2$ term in the second line of \WKBIII\
to match the units used in \COOT.  
Table 1 compares the WKB expressions with the numerical results found in
\COOT\ and we see that the agreement is very close.\foot{In \COOT\ a WKB 
expression was given with a numerical factor of $6$.}

\goodbreak
\vskip.5in
{\vbox{\ninepoint{
$$
\vbox{\offinterlineskip\tabskip=0pt
\halign{\strut\vrule#
&\hfil~$#$
&\vrule#&~
#&\hfil~$#$
&\vrule#&~
#&\hfil~$#$
&\vrule#&~
\hfil ~$#$~
&\vrule#
\cr
\noalign{\hrule}
& m &&& WKB~&& Numerical&& \cr
\noalign{\hrule}
& 1 &&  & 11.4843~ &&11.5877&& \cr
& 2 &&  & 34.453 ~ &&34.5270&& \cr
& 3 &&  & 68.906 ~ &&69.9750&& \cr
& 4 &&  & 114.853~ &&114.9104&& \cr
& 5 &&  & 172.265~ &&172.3312&& \cr
& 6 &&  & 241.171~ &&241.2366&& \cr
& 7 &&  & 321.561~ &&321.6265&& \cr
& 8 &&  & 413.436~ &&413.5009&& \cr
\noalign{\hrule}}
\hrule}$$
\vskip-7pt
\noindent
{\bf Table 1}: Comparison of $0^{++}$ glueball masses squared in units of 
$\pi^2T^2$.  The WKB approximation is very close to the numerical results
in \COOT, with a small difference approaching $0.064\pi^2T^2$ for large
$m$.
}
\vskip7pt}}

In the case of $QCD_4$, we have $n=2$ and hence the WKB relation
\eqn\WKBIV{
M~=~4\pi^{3/2}{\Gamma\left({2\over 3}\right)\over
\Gamma\left({1\over 6}\right)}T \sqrt{m(m+2)}~+~{\rm O}(m^{-1})\qquad
m\ge1
.}
Table 2 compares the WKB results in \WKBIV\
to the numerical results of \HashOz.  Again, we find that the WKB result
quickly approaches the numerical eigenvalues.
\goodbreak
\vskip7pt
{\vbox{\ninepoint{
$$
\vbox{\offinterlineskip\tabskip=0pt
\halign{\strut\vrule#
&\hfil~$#$
&\vrule#&~
#&\hfil~$#$
&\vrule#&~
#&\hfil~$#$
&\vrule#&~
\hfil ~$#$~
&\vrule#
\cr
\noalign{\hrule}
& m &&& WKB~&& Numerical&& \cr
\noalign{\hrule}
& 1 &&  & 9.39~ &&9.85&& \cr
& 2 &&  & 15.3 ~ &&15.6&& \cr
& 3 &&  & 21.0 ~ &&21.2&& \cr
& 4 &&  & 26.5~ &&26.7&& \cr
& 5 &&  & 32.1~ &&32.2&& \cr
& 6 &&  & 37.6~ &&37.7&& \cr
\noalign{\hrule}}
\hrule}$$
\vskip-7pt
\noindent
{\bf Table 2}: Comparison of $0^{++}$ glueball masses for $QCD_4$ in units of 
$T$.  The WKB approximation should approach the numerical result as a function
of $1/m$.
}
\vskip7pt}}

We conclude this section by examining the behavior of the mass spectrum
in the limit $n\to\infty$.  Taking this limit we approach
$p=5$, corresponding to six dimensional euclidean QCD with a compactified
direction.  In the large $n$ limit, the mass equation in \WKBfin\ 
reduces to
\eqn\WKBln{
M^2~=~4\pi^4T^2~\left(C~+{m\over n}\right)\qquad\qquad m\ge 1,
}
where $C$ is a constant to be determined.  Thus, it appears that for
nonzero $C$ there is
a finite gap in the spectrum, but above this gap the spectrum is
continuous. We can see this more clearly by taking $n\to\infty$ limit in
\diffeq.  In this case \diffeq\ reduces to Legendre's equation and so
the solutions that are regular at $x=1$ are $P_\ell(2x-1)$ where 
$\la=-\ell(\ell+1)$.  If $\la\le 1/4$ then $P_\ell(2x-1)$ is not normalizable
at infinity.  If $\la>1/4$, then $P_\ell(2x-1)$ is plane wave normalizable.
Therefore, we find that the constant in \WKBln\ is $C=\pi^{-2}$ and thus
there is a gap.  This unusual behavior for the six dimensional theory
is probably related to its nonlocal nature\refs{\ABKS,\PP}.

\newsec{Glueball masses for generalized supergravity backgrounds}

In the previous section we have seen that the analytic WKB expressions
give accurate results for the eigenvalues of the dilaton wave equation.
This strengthens are confidence in the procedure, and encourages us to
use it in more general situations.

In this section we discuss the WKB approximation for $0^{++}$ glueballs
in more general supergravity backgrounds.  We will argue that that there is 
universality in the spectra which
only depends on the dimensionality of the $AdS$ space at infinity and the
singularity structure at the horizon.  Using results derived here we
can find WKB approximations for $0^{-+}$ glueballs
in $QCD_4$ and $0^{--}$ glueballs in $QCD_3$.  We can also
explain the stability of the $0 ^{++}$ spectrum for models coming from 
rotating
branes \CORT, and the change in the spectrum for the $0^{-+}$ glueballs
in these same models.  

The only assumptions that we  make are that there are angular independent
solutions to the dilaton wave equation in \dileqm\ and that for large $U$
the metric approaches an $AdS$ solution.  In this case, we
can reduce \dileqm\ to
\eqn\dilgen{
\partial_U\left(f(U)\partial_U\phi\right)~+~M^2 h(U)~=~0,
}
where
\eqn\fhdef{
f(U)~=~\sqrt{g}e^{-2\phi}g^{UU},\qquad\qquad h(U)~=~\sqrt{g}e^{-2\phi}g^{xx}.
}
Let us suppose that there is a $U_0$ such that near $U=U_0$,
\eqn\fhsing{
f(U)~\sim~ (U-U_0)^s\qquad\qquad h(U)~\sim~(U-U_0)^q.
}
The assumption that asymptotically the solution is $AdS$ implies that
$f(U)\sim U^{8-p}$ and $h(U)\sim U$ if $U>>U_0$. 

Let us define $e^y=U-U_0$, $\tf(y)=f(e^y+U_0)e^{sy}$ and
$\th(y)=h(e^y+U_0)e^{qy}$.  If we let $\phi={e^{(1-s)y/2}\over\sqrt{\tf}}\psi$,
then \dilgen\ becomes
\eqn\ndgen{
\partial_y^2\psi~+~V(y)\psi~=~0,
}
where 
\eqn\Veq{
V(y)~=~M^2e^{(q+2-s)y}{\th(y)\over\tf(y)}~+~{1\over4}\left({\partial_y
\left(e^{(s-1)y}\tf(y)\right)\over e^{(s-1)y}\tf(y)}\right)^2
~-~\ha {\partial^2_y\left(e^{(s-1)y}\tf(y)\right)\over e^{(s-1)y}\tf(y)}.
}
For large negative and positive $y$ we have
\eqn\Vasymp{\eqalign{
V(y)~&\approx~C_1M^2e^{(q+2-s)y}~-~
{1\over4}(s-1)^2\qquad\qquad y<<0\cr
V(y)~&\approx~C_2M^2e^{(p-5)y}~-~
{1\over4}(7-p)^2\qquad\qquad y>>0,
}}
where $C_1$ and $C_2$ are unimportant constants for this discussion.  
Hence, we find
two turning points which for large enough $M$ can be approximated as
\eqn\ytp{\eqalign{
y_1~&\approx~-~{1\over q+s-2}\log\left({4C_1M^2\over(s-1)^2}\right)\cr
y_2~&\approx~{1\over 5-p}\log\left({4C_2M^2\over(7-p)^2}\right).
}}
The WKB approximation is then
\eqn\WKBgen{
\left(m+\half\right)\pi~=~\int_{y_1}^{y_2}dy\sqrt{V(y)}.
}

The leading order contribution to the integral in \WKBgen\ is
\eqn\WKBlo{
M\xi~=~M\int_{-\infty}^{+\infty}dy e^y\sqrt{e^{qy}\th(y)\over e^{sy}\tf(y)}
~=~M\int_{U_0}^\infty dU \sqrt{g^{xx}\over g^{UU}},
}
where we have used \fhdef.  It is clear that \WKBlo\ sets the 
inverse mass scale for the glueballs.  We can compare this to the scale
coming from the heavy quark potentials.  In this latter case,
the string tension along the brane, as a function of the energy scale $U$ is 
\eqn\strten{
\s(U)~={1\over2\pi}g_{xx}(U)
}
and at large quark separation the string tension 
approaches ${1\over2\pi}g_{xx}(U_0)$.
Hence we can rewrite \WKBlo\ as
\eqn\WKBss{
M\xi~=~M\int_{U_0}^\infty dU\sqrt{g_{UU}}(2\pi\s(U))^{-1/2}.
}
In other words, the inverse mass scale is a one-loop integral of the
square root of the inverse tension integrated over all energy scales
with a measure $\sqrt{g_{UU}}$.

Let us now consider the next to leading order corrections.  The computations
are similar to those in \corrone\ and \corrtwo.  Using \Vasymp,
the correction coming
from the turning point at $y=y_2$ is
\eqn\corrga{\eqalign{
&-\sqrt{M^2C_2}\int_{y_2}^\infty dy\ e^{(p-5)y/2}~+\cr
&\int^{y_2}_{-\infty}dy\left(\sqrt{M^2C_2e^{(p-5)y}-{1\over4}(7-p)^2}-
\sqrt{M^2C_2}e^{(p-5)y/2}\right)
~=~-\left({7-p\over5-p}\right){\pi\over2},
}}
while the correction coming from the turning point at $y=y_1$ is
\eqn\corrgb{\eqalign{
&-\sqrt{M^2C_1}\int^{y_2}_{-\infty} dy e^{(q+2-s)y/2}~+\cr
&\int_{y_2}^{+\infty}dy\left(\sqrt{M^2C_1e^{(q+2-s)y}-{1\over4}(s-1)^2}-
\sqrt{M^2C_2}e^{(q+2-s)y/2}\right)
~=~-{|s-1|\over q+2-s}{\pi\over2}. 
}}

Putting everything together, we find that the WKB masses are
\eqn\WKBmass{
M^2~=~{\pi^2\over\xi^2}~m\left(m+{2\over5-p}+{|s-1|\over q+2-s}\right)~+~
{\rm O}\left(m^0\right)\qquad\qquad m\ge1.
}
By assuming that the supergravity solution is asymptotically $AdS$, we
have chosen a particular singularity structure for the point at spatial
infinity.   The
basic arguments used here are still applicable even if the solution is not
$AdS$, so long as the singularity structure at infinity is known.  

We now consider some of the examples discussed in the recent literature.
For the case of rotating nonextremal 
D4 branes considered in \refs{\Russo,\CORT}, the
dilaton wave equation reduces to
\eqn\dilspin{
\partial_u\left[u\left(u^6-(4gN)^2a^4u^2-u_T^6\right)\partial_u\phi\right]
+4gNM^2u^3\phi~=~0,
} 
where $a$ parameterizes the angular momentum, and we have replaced $U$ with
$u^2=U$, to match the form of the equation in \CORT.  If $a=0$ this reduces
to the nonrotating D4 brane equation in \dileqmr.  The horizon occurs at
$u=u_0$ with
\eqn\horeq{
u_0^6-a^4u_0^2-u_T^6~=~0.
}
Hence, it is clear that \dilspin\ has the form of \dilgen\ and \fhsing,
with $s=1,q=0$, for {\it all} values of $a$.  Therefore,  we find that the WKB
expression for the masses is
\eqn\WKBgen{
M^2~=~m(m+2){\pi^2\over4gN}\left[\int_{u_0}^\infty du
{u\over\sqrt{u^6-(4gN)^2a^4u^2-u_T^6}}\right]^{-2}\qquad m\ge1.
}
Since the singularity structure does not change when $a$ is varied, we see
that the WKB mass ratios do not change either.
Thus, we see the reason for the stability of the glueball masses observed in
\CORT.  This might also explain
why supergravity glueball results are reasonably close to lattice results.

We can compute the integral in \WKBgen\ exactly in the two limits $a=0,\infty$.
The result for $a=0$ is in \WKBIV.  For large $a$ we find 
\eqn\WKBla{
M^2~=~{8\over\pi}\left(\Gamma\left({3\over4}\right)\right)^4 a^2m(m+2)
\qquad\qquad m\ge1.
}

The next examples are the $0^{-+}$ glueball masses for the rotating 
nonextremal D4 branes.  In this case, the equation of motion for one of
the angular components of the R-R 1-form field is \CORT
\eqn\eqmrra{
\partial_u\left[u^3(u^4-(4gN)^2a^4)
\partial_u \chi\right]+4gNM^2{u^5(u^4-(4gN)^2a^4)\over
u^6-(4gN)^2a^4u^2-u_T^6}\chi~=~0.
}
This equation also has the same form as \dilgen, but with $s=0$ and $q=-1$
for generic $a$.
Therefore, using \WKBmass we find
\eqn\WKBrra{
M^2~=~m(m+3){\pi^2\over4gN}\left[\int_{u_0}^\infty du
{u\over\sqrt{u^6-(4gN)^2a^4u^2-u_T^6}}\right]^{-2}\qquad m\ge1,
}
for the $0^{-+}$ masses.  In the $a=0$ limit this reduces to
\eqn\WKBIVrra{
M~=~4\pi^{3/2}{\Gamma\left({2\over 3}\right)\over
\Gamma\left({1\over 6}\right)} T\sqrt{m(m+3)}\qquad
m\ge1
.}
Table 3 shows a comparison of the WKB masses to the numerical results in
\HashOz.  The WKB result for the mass ratio between the lowest 
level  $0^{-+}$ and $0^{++}$ states is $M_{-+}/M_{++}=2/\sqrt{3}\approx 1.155$
which is reasonably close to the numerical result.  In fact the difference
between the WKB and numerical results is smaller than present day
lattice errors.
\goodbreak
\vskip7pt
{\vbox{\ninepoint{
$$
\vbox{\offinterlineskip\tabskip=0pt
\halign{\strut\vrule#
&\hfil~$#$
&\vrule#&~
#&\hfil~$#$
&\vrule#&~
#&\hfil~$#$
&\vrule#&~
\hfil ~$#$~
&\vrule#
\cr
\noalign{\hrule}
& m &&& WKB~&& Numerical&& \cr
\noalign{\hrule}
& 1 &&  & 10.8 ~ &&11.8&& \cr
& 2 &&  & 17.1 ~ &&17.8&& \cr
& 3 &&  & 23.0 ~ &&23.5&& \cr
& 4 &&  & 28.7~ &&29.1&& \cr
& 5 &&  & 34.3~ &&34.6&& \cr
& 6 &&  & 39.8~ &&40.1&& \cr
\noalign{\hrule}}
\hrule}$$
\vskip-7pt
\noindent
{\bf Table 3}: Comparison of $0^{-+}$ glueball masses for $QCD_4$ in units of 
$T$.  
}
\vskip7pt}}

However, in the limit that $a\to\infty$, the WKB structure will change for
the $0^{-+}$ states.  This is because the singularity structure of
\eqmrra\ changes.  In fact, in the large $a$ limit we end up with the
same equation as the dilaton.  Hence we find the same WKB masses.  The
only difference is that we have to discard the lowest eigenvalue \CORT.
Hence the WKB mass ratio in the large $a$ limit is 
${M_{-+}\over M_{++}}=\sqrt{8/3}\approx 1.63$.  This is again close to
the numerical result of $1.59$ and it is also close to
the lattice result of $1.61\pm.19$.

Our final example is the WKB spectrum for the $O^{--}$ glueballs in $QCD_3$.
After a rescaling of the NS-NS 2 form field, the relevant component satisfies
the wave equation \COOT
\eqn\zmmeq{
\partial_U\left[U^5(U^4-U_T^4)\partial_U\chi\right]+gNM^2U^5\chi~=~0.
}
This does not have quite the same form as \dilgen\ because of the extra
$U^4$ term, but the WKB analysis is almost identical and results in
the spectrum
\eqn\WKBIIIa{
M~=~\sqrt{8\pi} \left(\Gamma\left({3\over 4}\right)\right)^2 T\sqrt{m(m+3)}
\qquad m\ge1.
}
Table 4 is a comparison of the WKB and numerical results \KJMN.  Again,
we find close agreement.  Finally, the WKB mass ratio for the lightest
states with $0^{--}$ and $0^{++}$ quantum numbers is 
${M_{--}\over M_{++}}=\sqrt{2}$.

\goodbreak
\vskip.5in
{\vbox{\ninepoint{
$$
\vbox{\offinterlineskip\tabskip=0pt
\halign{\strut\vrule#
&\hfil~$#$
&\vrule#&~
#&\hfil~$#$
&\vrule#&~
#&\hfil~$#$
&\vrule#&~
\hfil ~$#$~
&\vrule#
\cr
\noalign{\hrule}
& m &&& WKB~&& Numerical&& \cr
\noalign{\hrule}
& 1 &&  & 4.79 ~ &&5.11&& \cr
& 2 &&  & 7.58 ~ &&7.82&& \cr
& 3 &&  & 10.17 ~ &&10.36&& \cr
& 4 &&  & 12.68~ &&12.84&& \cr
& 5 &&  & 15.16~ &&15.29&& \cr
& 6 &&  & 17.61~ &&17.73&& \cr
& 7 &&  & 20.05~ &&20.15&& \cr
& 8 &&  & 22.48~ &&22.57&& \cr
\noalign{\hrule}}
\hrule}$$
\vskip-7pt
\noindent
{\bf Table 4}: Comparison of $0^{--}$ glueball masses  in units of 
$\pi T$. 
}
\vskip7pt}}

\newsec{The type $0$ nonsupersymmetric model.}

In this section we make some general statements about the type $0$ model 
\refs{\DH,\SW}
recently discussed in \KT.  We find asymptotic solutions for the lowest
order supergravity approximation.  We find a running of the effective 
coupling, but no confining behavior and no normalizable glueball solutions.

The type $0$ model has a closed string tachyon,
no fermions and a doubled set of
 R-R fields.  In particular, there is no
longer a self dual constraint on the 5-form field strength.
Since the number of R-R fields is doubled, so are the number of D brane
types\BG.  Hence one can have D3 branes that are electric instead of
dyonic.  If we have $N$ parallel electric D3 branes, then the low energy
effective action on the branes is thought to be $SU(N)$ QCD with adjoint
scalar fields, but no fermions.  
Hence, there is no supersymmetry and the
coupling will run.  There is no open string tachyon \BG, so
there is no tachyon in this QCD model.

The authors in \KT\ argued that the {\it closed} string tachyon can get
an expectation value, and that its mass squared gets a positive shift
from the background 5 form flux.  The background tachyon field acts as
a source for the dilaton, so the dilaton is no longer constant.  One then
makes the following ansatz for the metric \KT
\eqn\typezm{
ds^2~=~e^{\half\phi}\left(e^{\half\xi-5\eta}d\rho^2~+~e^{-\half\xi}dx_{||}^2
~+~e^{\half\xi-\eta}d\Omega_5^2\right),
}
where $\phi$, $\xi$ and $\eta$ are functions of $\rho$ only.  The equations
of motion then reduce to a Toda like system with an action 
\KT\foot{We are using units where $\a'=1$.}
\eqn\toda{\eqalign{
S~=&~\int d\rho\left[\half \dot\phi^2+\half\dot\xi^2+{1\over4}\dot T^2
-5\dot\eta^2-
V(\phi,\xi,\eta,T)\right]\cr
V(\phi,\xi,\eta,T)~=&~\half T^2e^{\half\phi+\half\xi-5\eta}+20e^{-4\eta}-
Q^2f^{-1}(T)e^{-2\xi},
}}
and a constraint
\eqn\constraint{
\half \dot\phi^2+\half\dot\xi^2+{1\over4}\dot T^2-5\dot\eta^2+
V(\phi,\xi,\eta)~=~0.
}
 $Q$ is the total D3 brane charge which is proportional to $N$, $T$ is the
tachyon field and $f(T)$ is a function given by \KT
\eqn\fT{
f(T)~=~1+T+\half T^2~+~{\rm O}(T^3).
}
For large $Q$, the tachyon expectation value is determined by setting
 $f'(T)=0$.  As a first approximation, we may assume that the tachyon is
constant as a function of $\rho$. 

If $T=0$ then the solution reduces to the ${\cal N}=4$ solution.  
When $T$ is
nonzero, then all fields are coupled and there is no known analytic solution.
However, we can attempt to find approximate solutions that are valid in
the {\it UV} and {\it IR} regions.  In the {\it UV}, 
we expect the dilaton field to
be relatively constant, at least compared with $\xi$ and $\eta$.  Assuming
that $\phi$ is constant and thus ignoring its kinetic term, the equations for
$\xi$ and $\eta$ can be solved exactly, at least in the near horizon limit.
In this case we find 
\eqn\appsol{\eqalign{
&e^{\xi}~=~C_1\rho \qquad\qquad e^\eta~=~C_2\rho^{1/2}\cr
&{1\over4} T^2e^{\ha\phi}{C_1^{1/2}\over C_2^5}~+~{2Q^2\over C_1^2f(T)}
~-~1~=~0\cr
&{5\over2}T^2e^{\ha\phi}{C_1^{1/2}\over C_2^5}~+~{80\over C_2^4}~-~5~=~0.
}}
One can easily check that this satisfies the constraint equation in 
\constraint. If we plug this back into the metric, we find that the solution
is still $AdS_5\times S_5$, but the curvatures
 of the two spaces no longer match,
 $S_5$ now has smaller curvature then $AdS_5$.
In this case the Ricci scalar for the total space is proportional to
\eqn\ricci{
R~\sim~ T^2 e^{\half\phi}.
}

Using the  $\xi$ and $\eta$ solutions as inputs, we can go back and find
an approximate solution for
 $\phi$ in terms of $\rho$. Using the ansatz
 $e^{\half\phi}=C_0(\log(\rho/\rho_0))^\a$, and
plugging this into the equation of motion for $\phi$
\eqn\phieqm{
\ddot\phi~+~{1\over4} T^2 e^{\half\phi+\half\xi-5\eta}~=0,
}
we find that the ansatz is a leading order solution if $\a=-1$ and
 $C_0=-8C_2^5/(T^2\sqrt{C_1})$.  $\rho_0$ is an integration constant and we 
assume that $\rho_0>>1$ in order that the gauge theory length scale is
much greater than the string scale.  Setting $\rho=u^{-4}$, and using the
lowest order solutions for $C_1$ and $C_2$ from \appsol, we learn that the
leading order behavior for the coupling is
\eqn\couplo{
e^{-\phi}~=~{1\over g_{YM}^2}~=~{QT^4\left(\log {u\over u_0}\right)^2\over 
4096\sqrt{2f(T)}}.
}
Thus we find a running coupling, but instead of a linear log dependence,
the coupling runs with a log squared!    One can easily check
that to leading order in $1/\log u$, the constraint equation is still
satisfied.  We can also estimate the range of validity for this solution.
Computing the leading order corrections to $C_1$ and $C_2$, one finds that
\eqn\corrcs{
C_1~=~{2Q\over\sqrt{2f(T)}}\left(1+{1\over 4\log {u\over u_0}}\right)\qquad
C_2~=~2\left(1+{1\over 4\log {u\over u_0}}\right).
}
We can also compare the terms in the potential that depend on the tachyon.
Since 
\eqn\tachterms{
{1\over4} T^2 e^{\half\phi+\half\xi-5\eta}\sim {u^8\over 2\log {u\over u_0}}
\qquad
Q^2f^{-1}(T)e^{-2\xi}\sim {u^8\over2},
}
our solution with  constant $T$ and $f'(T)=0$ is valid
 so long as $\log(u/u_0)>>1$.

The metric in the large $u$ limit is 
\eqn\metlu{
ds^2~=~{32\over T^2\log {u\over u_0}}\left({du^2\over u^2}+{\sqrt{2f(T)}
\over2Q}
\left(1+{1\over\log {u\over u_0}}\right)u^2dx_{||}^2+
\left(1+{1\over\log {u\over u_0}}\right)
d\Omega_5^2\right).
}
Hence we can  trust the supergravity solution only if $T<<1$, since
 $\log (u/u_0)>>1$.  However,
it is clear that $T\sim1$ if $f'(T)=0$, hence one should not expect
 the supergravity result
to be particularly trustworthy.  Indeed, we have found that while
the supergravity computation results in a  coupling running to zero, 
it runs with the wrong power of $\log u$. 

Nevertheless, the effective coupling between a heavy quark and its antiquark
{\it does} appear with the expected log dependence.  Using the Wilson line
computation of \refs{\RY,\JMaldb}, one finds that the quark potential is
given by
\eqn\qp{
V~\approx~-~{256\ \pi^3\over  \Gamma({1\over4})^4T^2L\log(L_0/L)}
\qquad\qquad L<<L_0}
where 
 we have plugged $R^2={32\over T^2\log (u/u_0)}$ into the expressions derived
in \refs{\RY,\JMaldb}. $L_0$ is some length that can be adjusted to
be much longer than the string scale.
 Recall that the
${\cal N}=4$ potential comes
with a coefficient $\sqrt{gN}$.  It is this coefficient, and  not
 the 't~Hooft-Polyakov tension $gN$, that plays the role of
$1/\a'$ for the supergravity models.  For a string theory with 
extrinsic curvature, the string
tension has a $(\log u)^{-1}$ dependence\refs{\Polyakov,\Horava}.  
Hence our result is in line with  rigid string results, so long 
as one remembers that the tension is $\sqrt{gN}$.


Even though we cannot really trust the supergravity solution for large $u$, we
might be able to trust it for small $u$.  However, here we will see that the
situation is even worse.  In particular, we will find that the dilaton
wave equation in the lowest order supergravity background has no normalizable 
glueball solutions.  Moreover, the heavy quark potential is {\it not} 
confining.

In order to study the {\it IR} behavior, 
let us follow the suggestion of \KT\ and search for solutions for the toda 
system assuming
that the second term in $V$ is small compared to the other terms.
This corresponds to a small curvature for the  $S_5$.  Dropping
this second term in $V$ and assuming a constant $T$, one can now
find an exact solution to the equations of
motion that satisfies the constraint.  The solution is
\eqn\lrhosol{
e^\phi~=~{C_0^2}\rho^{5/9}\qquad e^\eta~=~C_2\rho^{5/9}
\qquad e^{\xi}~=~{3Q\over\sqrt{2f(T)}}\rho,
}
with the relation
\eqn\conrel{
20(2f(T))^{1/4}C_2^5-9T^2\sqrt{3Q}C_0~=~0.
}
 Comparing all terms in $V$, one has
 $e^{\half\phi+\half\xi-5\eta}\sim \rho^{-2}$,
$e^{-2\xi}\sim \rho^{-2}$, but $e^{-4\eta}\sim \rho^{-20/9}$. Hence
this solution is valid for large $\rho$.  From \lrhosol, the 
coupling blows up as $\rho\to\infty$ and after substituting $\rho=1/u^4$
the metric is
\eqn\metsmrho{
ds^2~=~{20\over9 T^2}\left(16{du^2\over u^2}+{C_2^5\sqrt{2f(T)}\over 3Q}
u^{8/9}
dx_{||}^2+C_2^5u^{-8/9}d\Omega_5^2\right).
}
$C_2$ remains as a leftover integration constant and is ultimately
determined by matching to the {\it UV} solution. 

In the small $u$ limit, the   $T$ dependent terms in the potential
are now comparable, so the tachyon expectation value is no longer at
$f'(T)=0$.  Instead, plugging in the solution of \lrhosol\ and \conrel\ into
the tachyon equation of motion, one finds that $\ddot T=0$ if 
\eqn\Trel{
10f(T)~+~Tf'(T)~=~0.
}
Using the function in \fT\ for $f(T)$, one finds that there are no real
solutions of \Trel\ for $T$. However, if we were to include the cubic term
in $f(T)$ or include the quartic
term  $c_1T^4$ in the action, then a real solution would
exist.  

 As in the {\it UV}, the curvature in the {\it IR} is
 small if $T<<1$.   However, we expect solutions to \Trel\ to be
 $T\sim1$.  The curvature is now at the string scale, so we cannot
 truely  trust the supergravity solution in this limit either.   

We can attempt to find glueball solutions.  Using
 the asymptotic expressions  in the dilaton equation of motion
we find the equations
\eqn\diltyz{\eqalign{
&\partial_u\left[u^5\partial_u \phi\right]~+~{2Q\ M^2\over\sqrt{2f(T)} }u
\phi~=~0
\qquad u>>u_0\cr
&\partial_u\left[u^5\partial_u \phi\right]~+~{48Q\ M^2\over C_2^5\sqrt{2f(T)}}
u^{21/9}\phi
~=~0\qquad u<<u_0
}}
For large $u$, we see that the dilaton equation of motion reduces to the
 ${\cal N}=4$ result.  It appears that the  only other singularity 
 is at $u=0$.
Using the arguments of the previous section, one learns from
\diltyz\ that this singularity
 has $s-q=5-21/9>2$, hence no glueball solutions
are possible.  

We can also easily see that the potential between the heavy quarks does
not confine for these solutions.  If we define a new variable $v$ such that 
\eqn\vrel{
v~=~{1\over9}\left({\sqrt{2f(T)}\over 3Q}\right)^{1/2}~u^{4/9},
}
then the metric in \metsmrho\ is
\eqn\metv{
ds^2~=~{180\over T^2}\left({dv^2\over v^2}~+~v^2dx_{||}^2~
+~{(C_2/3)^{10}\sqrt{2f(T)}\over Q^2v^2}d\Omega_5^2\right).
}
From this metric, we see that $R^2={180\over T^2}$, and so the heavy quark
potential is
\eqn\qpII{
V~\approx~-~{1440\ \pi^3\over  \Gamma({1\over4})^4T^2L}
\qquad\qquad L>>L_0.
}
Hence, the supergravity result implies that the effective heavy quark coupling
increases when going from the {\it UV} to the {\it IR}, but
 the quark potential does not develop
 a linear term and  remains coulombic.

We could certainly generate a linear quark potential by going to finite 
temperature since the supergravity background would now have a horizon at
finite $u$.  The down side of this scenario is that the the theory will
essentially be reduced to QCD in three dimensions. 
 
In the end one probably has to consider the  full $\s$-model in order
to get a linear quark potential and  normalizable
glueball solutions.
 At the very least, one could try
including the $\a'^3R^4$ terms in the action to see if this would qualitatively
change the behavior in the infrared.  Perhaps one could then combine the {\it
UV} results found here with {\it IR} results derived from the $\s$-model to say
something concrete about the WKB glueball masses.

\newsec{Conclusions}

We have seen from the WKB mass expressions that there is some
degree of universality for glueball mass ratios between different 
supergravity models.  In the examples where the ratios change, the behavior
can be attributed to a change of the singularity structure at the horizon.

The finite temperature models do not exhibit a running of the coupling in
the {\it UV}.   However, we have shown that the type 0 model has the
desired behavior.  Unfortunately, it does not appear to
be confining in the {\it IR}.   Hopefully confinement will appear in
the solution for the full $\s$-model.
Or  perhaps a model can be found that combines
the desired features of the nonsupersymmetric models discussed here.

\goodbreak
\vskip2.cm\centerline{\bf Acknowledgements}
\noindent

I would like to thank O. Bergman, E. Gimon, P. Horava and D. Minic
for discussions.
This work was supported in part
by funds provided by the DOE under grant number DE-FG03-92-ER40701.

\goodbreak

\listrefs
\end